Letter to *The Times*
'Innovative Science', Wednesday 23 September 2015 p26.

Sir, We write as senior scientists about a problem vital to the scientific enterprise and prosperity. Nowadays, funding is a lengthy and complex business. First, universities themselves must approve all proposals for submission. Funding agencies then subject those that survive to peer review, a process by which a few researchers, usually acting anonymously, assess a proposal's chances that it will achieve its goals, is the best value for money, is relevant to a national priority and will "impact" on a socio-economic problem. Only ~25% of proposals received by the funding agencies are funded. These protracted processes force researchers to exploit existing knowledge, severely discourage open-ended studies and are hugely time-consuming. They are also new: before ~1970, few researchers wrote proposals. Now they are virtually mandatory.

Globally, the university sector has expanded three or four-fold since ~1970. However, the $20^{th}$ century was dominated by discoveries made by ~500 Nobel Prize winners, almost all of whom began their work before ~1970. Almost all were academics exploring new concepts. Their work led to nuclear power; penicillin; lasers; magnetic resonance imaging and monoclonal antibodies. Most were younger than 40 when they made their discoveries and few, if any, were predicted.

The potential for discovery now is as great as ever, but we must find new ways of giving unconstrained support to the tiny number of scientists with radical agenda. Venture Research was British Petroleum's initiative for supporting such people led by one of us (DWB). It ran from 1980 to 1993, created at least 14 major discoveries from the 37 groups supported, all of whom, except possibly one, had been rejected by peer review. Its total cost, including BP and university overheads, was some £20 million over 13 years. Identifying people to lead such initiatives will be difficult but it must be done.

Donald W Braben*
University College London,
to whom correspondence should be addressed and the following, who also sign in a personal capacity:

John F Allen, University College London;

William Amos, **FRS**, University of Cambridge;

Richard Ball, University of Edinburgh;

Hagan Bayley, **FRS,** University of Oxford;

Tim Birkhead **FRS**, University of Sheffield;

Peter Cameron, Queen Mary, University of London;

Eleanor Campbell, University of Edinburgh;

Richard Cogdell **FRS**, University of Glasgow;

David Colquhoun **FRS,** University College London;

Steve Davies, University of Oxford;

Rod Dowler, Industry Forum, London;

Peter Edwards, **FRS**, University of Oxford;

Irene Engle, US Naval Academy, Annapolis;

Felipe Fernández-Armesto, University of Notre Dame;

Desmond Fitzgerald, Materia Medica;

Jon Frampton, University of Birmingham;

Dame Anne Glover, University of Aberdeen;

John Hall, University of Colorado, **Nobel Laureate**:

Pat Heslop-Harrison, University of Leicester;

Dudley Herschbach, Harvard University, **Nobel Laureate**;

Sui Huang, Institute for Systems Biology, Seattle;

H Jeff Kimble, Caltech, **US National Academy of Sciences;**

Sir Harry Kroto **FRS**, Florida State University, **Nobel Laureate**;

James Ladyman, University of Bristol;

Peter Lawrence **FRS**, University of Cambridge;

Mark Leake, University of York;

Armand Leroi, Imperial College London;

David Logan, University of Oxford;

Angus MacIntyre **FRS**, Queen Mary, University of London;

Julian Marchesi, Cardiff University and Imperial College London;

John Mattick **FAA**, Garvan Institute of Medical Research, Sydney;

Colin McInnes **FREng**, University of Glasgow;

Tom McLeish **FRS**, Durham University;

Graham Medley, London School of Hygiene and Tropical Medicine;

Thomas Miller, University of California at Riverside;

Randolph Nesse, Arizona State University;

Gerald Pollack, University of Washington;

Beatrice Pelloni, University of Reading;

Douglas Randall, **Member National Science Board**, University of Missouri;

David Ray, formerly Deputy Director BP Venture Research;

Sir Richard J Roberts **FRS**, New England Biolabs, **Nobel Laureate;**

Ken Seddon, Queen's University of Belfast;

Colin Self, University of Newcastle;

Harry Swinney, University of Texas, **US National Academy of Sciences**;

Chris Thomas **FRS**, University of York;

William Troy, University of Pittsburgh;

Robin Tucker, University of Lancaster;

Claudio Vita-Finzi, **FBA,** Natural History Museum;

David Wild, University of Warwick.